\documentclass[aps,prc,twocolumn,floatfix,showpacs,preprintnumbers,amsmath,amssymb,superscriptaddress]{revtex4}

\usepackage{graphicx,amsmath,color}
\usepackage{latexsym}

\usepackage[switch]{lineno}
\definecolor{lightred}{rgb}{1,0.5,0.5}
\definecolor{lightgreen}{rgb}{0.5,1,0.5}
\definecolor{darkgreen}{rgb}{0.5,0.9,0.5}

\begin{document}


\title{ History and Geography of  Light Pentaquark Searches:\\
Challenges and Pitfalls
}
\author{M.~Amaryan\\
Department of Physics,
Old Dominion University\\
Norfolk,  VA 23529,  USA}

\date{\today}
%

\begin{abstract}
In this paper I review the history and geography of the pentaquark searches and discuss  the current situation surrounding these searches performed at different facilities around the world.
The possibility of the existence of multiquark states like tetraquarks and pentaquarks was already predicted by
Gell-Mann~\cite{gell-mann} based on the Constituent Quark Model (CQM),  however more than half a century efforts in a wide range of experiments led to
controversial situation, when the fate of the light quark pentaquarks is almost decided to not exist. The  recent 
LHCb results ~\cite{LHCB} on the observation of the charm pentaquarks  in the invariant mass of  $pJ/\psi $  from the $\Lambda_b\to K^- p J/\psi$ decay created a new wave of excitement and rises the question about the existence of the  light pentaquarks.
The main question which still remains to be clarified is whether  already acquired evidences are sufficient to completely disregard the light pentaquarks and leave it out as an example of the  scientific curiosity or there are still rooms for further,  more dedicated efforts and scrupulous analyses to answer the question of the existence or non existence of the light pentaquarks made of $u, d$ and $s$ quarks.

\end{abstract}

\pacs{12.38.Aw, 13.60.Rj, 14.20.-c, 25.20.Lj}

\maketitle

\section{ Introduction}

In his fundamental paper  
~\cite{gell-mann},  Gell-Mann    anticipated the existence of multiquark states including  pentaquarks based on the Constituent Quark Model.  The existence of pentaquarks does not contradict any basic principles of the quantum field theory of strong interactions,  Quantum Chromodynamics (QCD) either.  However,  their spectra are hard to calculate due to the perturbative character of QCD.  However,  for a long period of time the search for pentaquarks before 2003 was sporadic until people started paying attention to the prediction of the chiral soliton model by Diakonov,  Petrov and Polyakov ~\cite{DIAKON}.
In this  paper the entire new family of the anti-decuplet pentaquarks was predicted.  The lightest   member of this multiplet, the $\Theta^+$,  with $uudd\bar s$ quark content decaying to $K^+n$ or $K^0p$ attracted experimentalists due to  a very narrow width of 15~MeV (or less) and  the mass of 1530~MeV accessible at many facilities around the world.  It was  naively expected to be relatively easy to  discover.  Alas,  as  will be  discuss below,  it was not to be the case after all.

\section{Positive Claims}
So, what happened in 2003? The LEPS Collaboration at  SPring-8 in Japan published the paper ~\cite{nakano} claiming an observation of $\Theta^+$ in a photoproduction experiment on the $^{12}C$ target.
The paper reported an  observation of a resonance structure in the invariant mass of $K^+n$ from the reaction $\gamma+n\to K^+K^-n$ with a mass of M($K^+n)$=$1.54\pm 0.01$GeV/c$^2$ with a width less than 25~MeV.  The statsical significance of the observed structure was reported to be 4.6$\sigma$.

The paper published by DIANA collaboration ~\cite{DIANA} was done independently using $K^+$ scattering in a  bubble chamber filled with a Xe.  As it was reported,  in the charge-exchange reaction $K^+ Xe \to K^0p Xe^{\prime}$ the spectrum of $K^0p$ shows a resonant enhancement with 
M=1539$\pm$~2~MeV/c$^2$ and $\Gamma \leq 9$~MeV/c$^2$.  The statistical significance of the observed enhancement was quoted to be near 4.4$\sigma$.
Both these experiments associate their observed peaks with the lightest member of the anti-decuplet predicted by chiral soliton model ~\cite{DIAKON}. 

The next in a row was a paper by the CLAS collaboration  ~\cite{CLASD}.   In the abstract of this paper it is written: "In the exclusive measurement of the reaction $\gamma d \to K^+ K^-pn$,  a narrow peak that can be attributed to an exotic baryon with strangeness $S = +1$
is seen in the $K^+n$ invariant mass spectrum". The mass of the peak  at $1.542\pm 0.005$~GeV/c$^2$ and FWHM width of 0.021~GeV/c$^2$ was reported. The statistical significance of the peak was estimated  to be $5.2\pm 0.6\sigma$.

The CLAS Collaboration searched for $\Theta^+$ also in the photoproduction reaction $\gamma p\to \pi^+ K^- K^+ n$ on a proton target ~\cite{CLASH}.  An  observation of the peak in the invariant mass  M$(K^+n)$=1.55$\pm $10~MeV with a FWHM of $\Gamma$= 26MeV/c$^2$ was reported.  The statistical significance of the signal was estimated at 7.8$\pm 1 \sigma$.

Consequently papers were published  by different collaborations ~\cite{SAPHIR,  HERMES,   ZEUS,  COSY,  SVD} all claiming an observation of $\Theta^+$ with the mass around $\sim$1540MeV and the width about 20~MeV or less.  The common feature of all experiments done in the different countries in the world,  in Japan,  Russia, United States, Germany,  Switzerland,  China at different facilities  with different beams and  energies was a low statistics,  which in some cases resulted in the observation of the peak and possibly overestimation of the statistical significance in some of them.

 \section{Negative Results}

Even besides the high statistics measurement of the CLAS,  which I will discuss below,  there were reports on non observation of $\Theta^+$ in some other experiments ~\cite{HERAB,  HyperCP,  BES,  ALEPH,  BABAR}.  In 2004 the CLAS collaboration at Jefferson Laboratory performed new dedicated photoproduction experiments both on a deuteron and hydrogen targets.
The main goal of these experiments was to check whether previous claims of the observation of $\Theta^+$ in a  low statistics experiments will be confirmed or not with about 10 times higher statistics. The first results on non observation at high statistics CLAS experiment in the reaction $\gamma d \to K^+K^-pn$ dismissing previous ~\cite{CLASD} results was presented in April 2005 during the APS Meeting in Tampa, Florida.  Results of this experiment were published in ~\cite{CLASg10}. Consequently the results of $\Theta^+$ search in high statistics experiment on the hydrogen target were  published  in ~\cite{CLAS1} with non observation of $\Theta^+$.

\section{Discussion}

Before making  a verdict  on the fate of $\Theta^+$ let us discuss what are the challenges and pitfalls of performed searches.  First of all let us mention that the results of ~\cite{nakano} and ~\cite{DIANA} experiments have not been refuted.  As the high statistics CLAS results had an undoubtfully the most significant impact on the current status of essentially non existence of $\Theta^+$ let us discuss them in more details.

The common feature of all experiments with primary beams with no strange content unavoidably lead to at least three particle final states.  In the photoproduction experiments searching for $\Theta^+$ pantaquark  the following reactions can be studied

\begin{eqnarray}
\gamma + d  \to K^+K^-pn\\
\gamma + p \to K^+\pi^+ K^-n\\
\gamma + p \to \bar {K^0} K^+n\\
\gamma + p \to \bar {K^0} K^0 p
\end{eqnarray}

\begin{figure}[htb!]
\includegraphics[ width=3.2in] {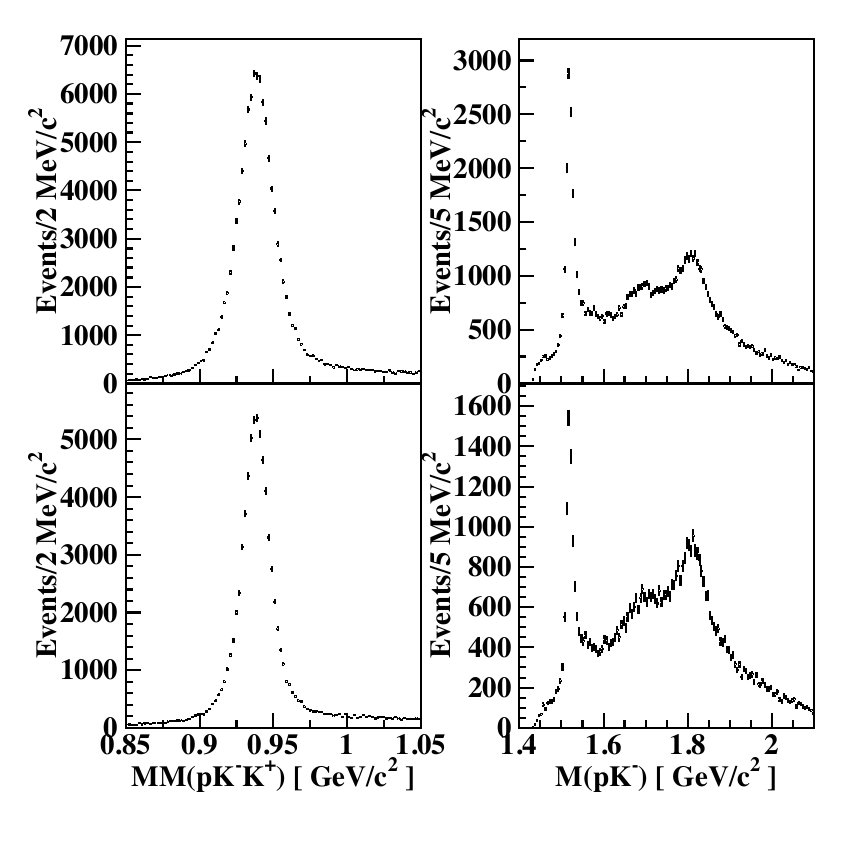}
\caption{(Color online) Missing mass of all charged particles (left) and the invariant mass $M(pK^-)$ (right panels). The data are shown for two magnetic fields of CLAS setup ~\cite{CLASg10}.}
\label{fig:m_kp}
\end{figure}

\begin{figure}[htb!]
\includegraphics[ width=3.2in] {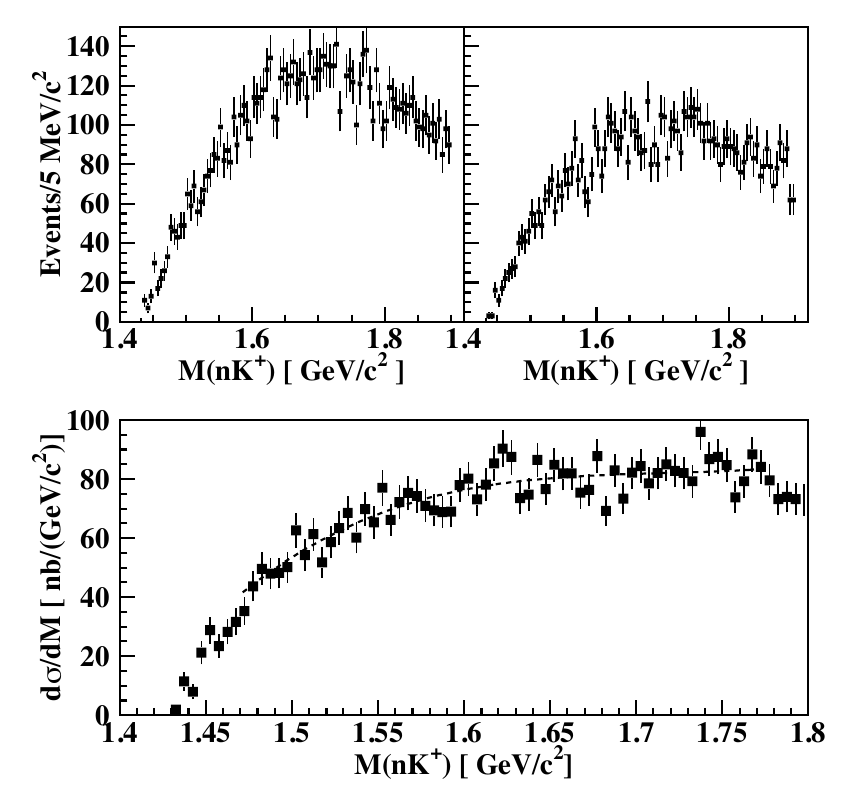}
\caption{(Color online) In the upper two panels the invariant mass $M(K^+n)$ distributions are plotted for two different magnetic fields,  while in the lower panel the photoproduction cross section is plotted as a function of  $M(K^+n)$ invariant mass, for details see  ~\cite{CLASg10}.}
\label{fig:m_kn}
\end{figure}

\begin{figure}[htb!]
\includegraphics[ width=3.2in] {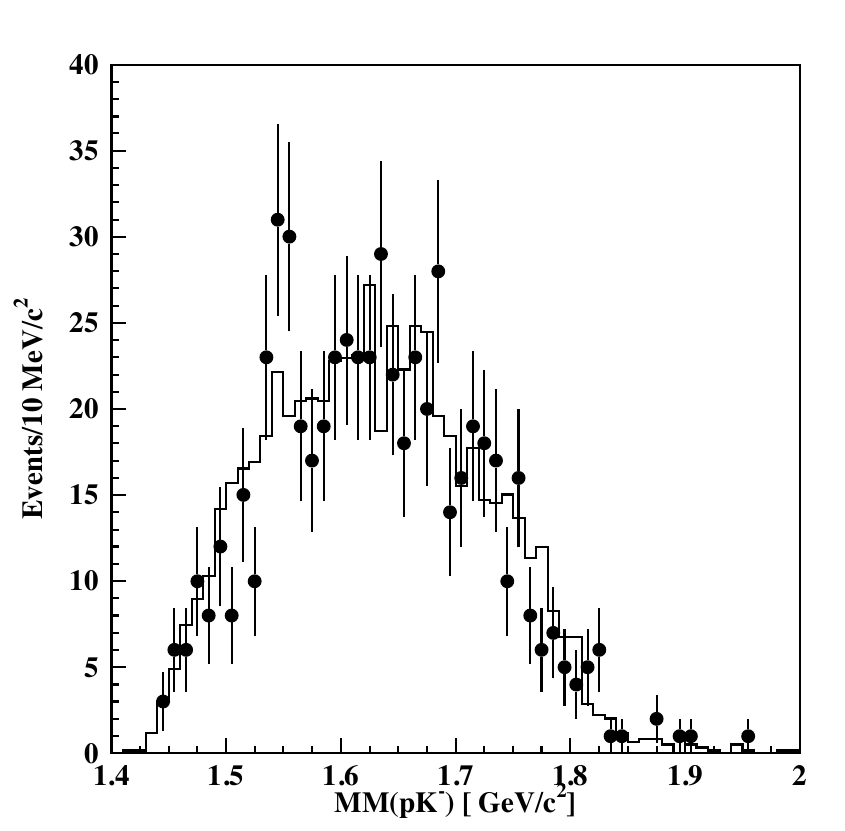}
\caption{(Color online) Comparison of the previously published ~\cite{CLASD} result with the high statstics experimental resul from ~\cite{CLASg10} normalized by 1/5.92.}
\label{fig:comp}
\end{figure}

The reaction of eq.1  studied in ~\cite{CLASD}  and ~\cite{CLASg10}. To avoid reflections, as  the $K^+K^-$ can make a $\phi$ meson,  a cut was imposed on  $M(K^+K^-)>$1.06~GeV,. On the other hand the invariant mass of $K^-p$  system is very reach making not only $\Lambda(1520)$ but also higher mass $\Lambda^*$'s and $\Sigma^*$'s,  however only events under the peak of $\Lambda(1520)$ were removed,  by a cut $1.495<M(K^-p)< 1. 545$~GeV/c$^2$ see Fig.~\ref{fig:m_kp}.  As one can see the overlapping events above 1.545~GeV/c$^2$ still remain in the invariant mass of $M(K^+n)$ see Fig.~\ref{fig:m_kn}. 
Finally the obtained high statistics distribution was compared with the previously published one in ~\cite{CLASD}, see Fig.~\ref{fig:comp}

What can be concluded from this comparison? That previous distribution which looked like having a structure around 1.54~GeV has not been reproduced.  The fact is that 
the claim of the observation of the structure with high significance in a previously published paper may have been ruled out if the log likelihood estimation of the significance would have been performed or Kolmogorov-Smirnov test for the signal+background or the background only hypotheses could have been used. 

\begin{figure}[htb!]
\includegraphics[ width=3.2in] {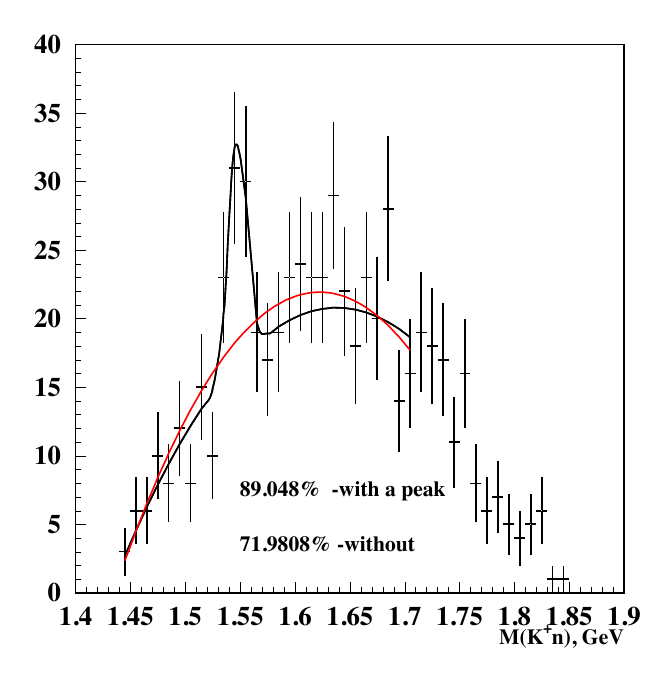}
\caption{(Color online) The Kolmogorov-Smirnov tests performed for the "peak hypothesis" and the "null hypothesis" show the confidence level of both assumptions.}
\label{fig:ks-test}
\end{figure}

In Fig.~\ref{fig:ks-test} results of such a test are presented.  As one can see  the goodness of the fit of the "peak hypothesis" is about $90\%$ while for the "null hypothesis" is 72$\%$,  which is much too high to be excluded.

The fact that there are so much phase space for the high mass $K^-p$ to contribute to the final plot of $K^+n$ invariant mass,  it is premature to conclude that the existence of $\Theta^+$
 is ruled out.  It may have been interesting to see  if other cuts like for example a cut on a $t$-Mandelstam, 
 could have suppressed higher mass $K^-p$ excited hyperons more than the possible pentaquark,  however such studies were not performed.
 
\begin{figure}[htb!]
\includegraphics[ width=3.2in] {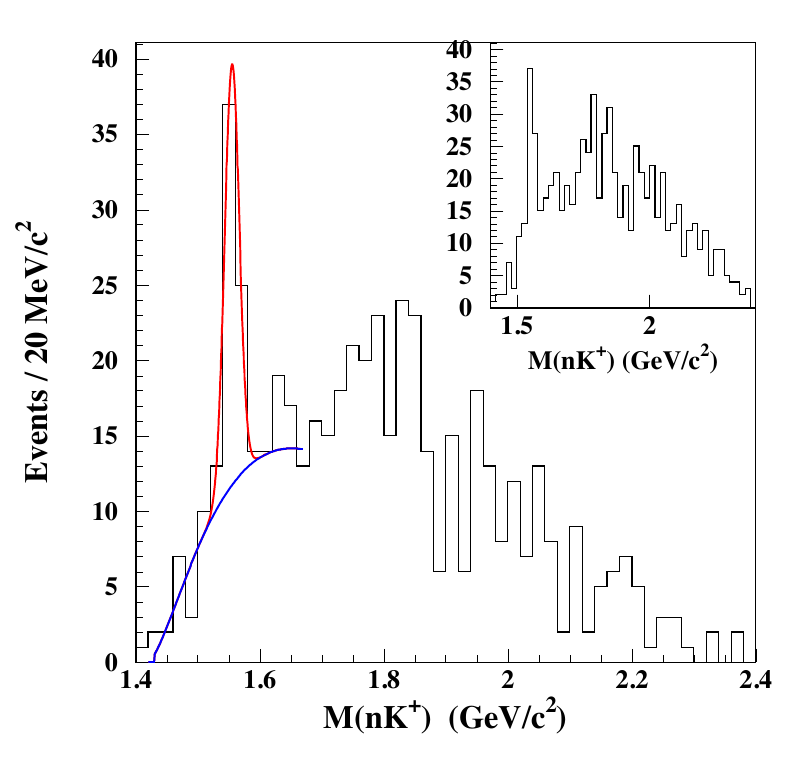}
\caption{(Color online) The $nK^+$ invariant mass spectrum. For details about cuts applied see ref.~\cite{CLASH}.}
\label{fig:kub}
\end{figure}

 The search for the $\Theta^+$ was also performed using reaction 2 above ~\cite{CLASH}.  
The final state in this reaction is even more complicated with four particles in the final state.
The authors of ~\cite{CLASH} removed the $\phi$ peak by the cut $M(K^+K^-) >1.06$~GeV/c$^2$. Here also the significance quoted to be $7.8\pm 1$ is clearly overestimated.  However, besides this there were not any discussion in the paper how much reflections could come from $K^*(890)\to K^-\pi^+$ and whether  the  reflections from the high mass $\Sigma^{*-}$ decaying to $K^-n$ can populate $K^+n$ spectrum.  The $K^+ n$ spectrum from ~\cite{CLASH}  is presented in Fig.~\ref{fig:kub}.

\begin{figure}[htb!]
\includegraphics[ width=3.2in] {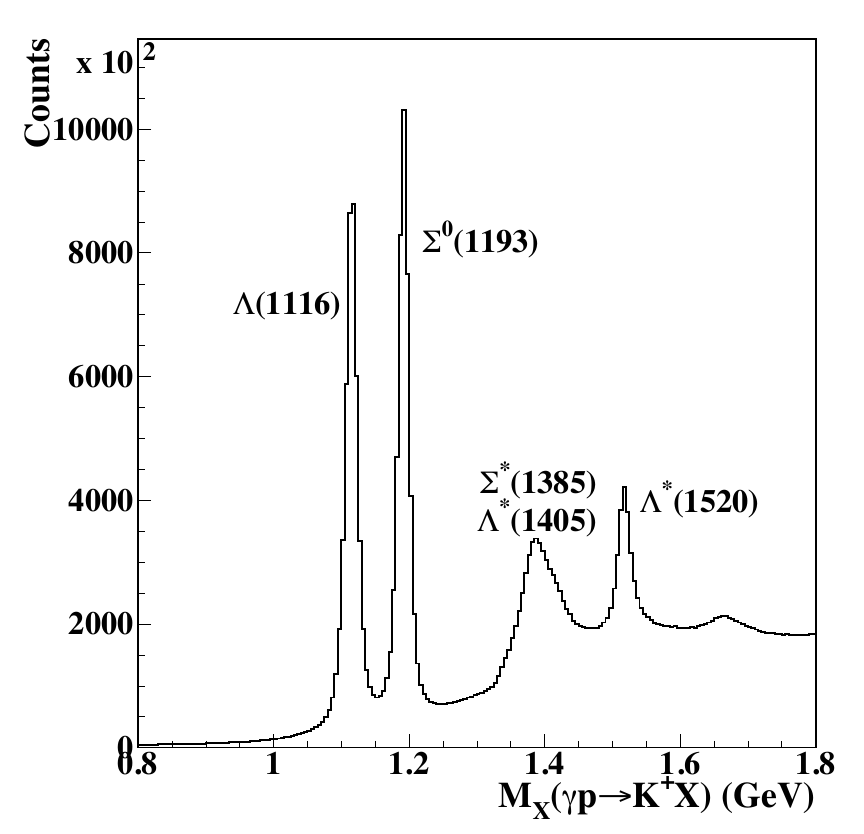}
\caption{(Color online) The missing mass of $K^+$.   See ref.~\cite{CLAS1} for more details.}
\label{fig:batt1}
\end{figure}

\begin{figure}[htb!]
\includegraphics[ width=3.2in] {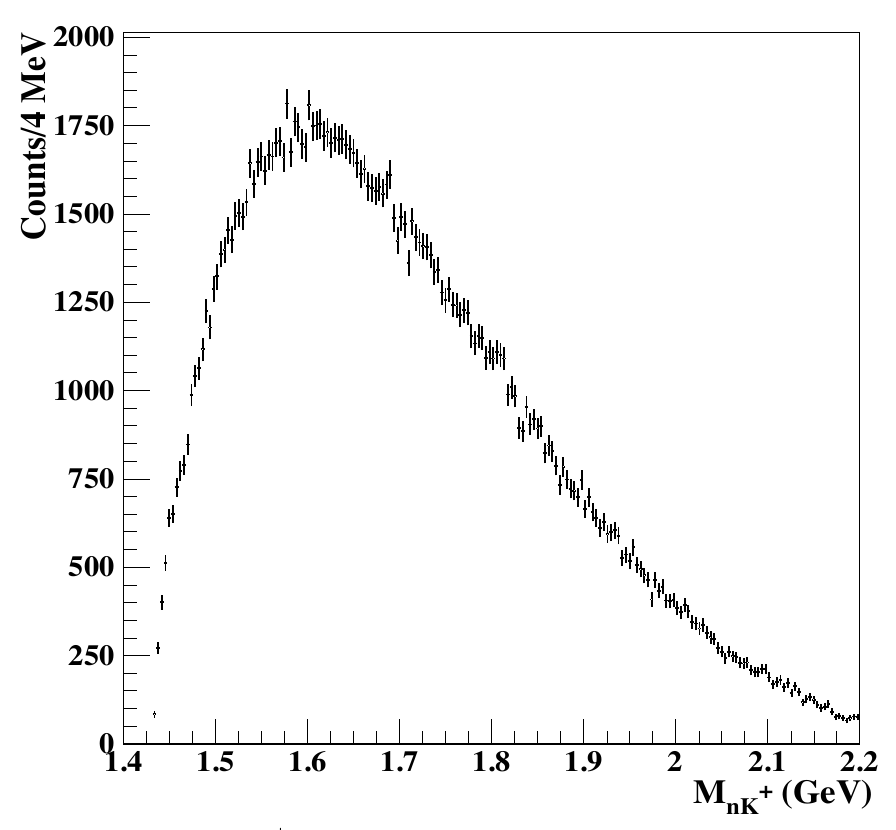}
\caption{(Color online) The  $nK^+$. invariant mass spectrum for the reaction $\gamma p \to \bar K^0 K^+n$  after all cuts.  For details see ~\cite{CLAS1}. }
\label{fig:batt2}
\end{figure}

The search for the $\Theta^+$ was performed also in ~\cite{CLAS1} using reactions 3 and 4 above on the proton.  In the reaction $\gamma + p \to \bar K^0 K^+n$ the main source of the reflections on the invariant mass of $K^+n$ system is due to the excited hyperon  states, in this case $\Lambda^*$'s as well as $\Sigma^{*0}$'s. The missing mass of $K^+$ is presented in Fig.~\ref{fig:batt1} from ref.~\cite{CLAS1}, where there are clearly other states above $\Lambda(1520)$. The final structureless figure is presented in Fig.~\ref{fig:batt2}.

In the next studied reaction was number 4 above,  $\gamma p \to K_S K_L  p$.  Here the $K_L$ was a missing particle, the $K_S$ was reconstructed in the invariant mass of oppositely charged pions. The possibility of $K_SK_L$ making a $\phi$ was cut out by selecting events with missing mass of the proton  above $\phi$,  with $MM(p)>1.04$~GeV and the possibility of $M(\pi- p$ to make a ground state $\Lambda$ was excluded by the cut $M(\pi p) > 1.13$~GeV.  However even in this case the fact that when searching for $\Theta^+$ in the missing mass of $K_S$ the excited $\Sigma^*$'s in the invariant mass of $p K_S$ can be reflected on the $M(pK_L)$, which is $M_X(K_S)$,  and similarly excited $\Sigma^*$'s contribution in the invariant mass $M(p K_S) $  was ignored and no measures have been taken to avoid such reflections.  
One has to mention that the family of excited hyperon states is not well established both theoretically and experimentally which may result in a large uncertainty when making a decision about reflections.

\begin{figure}[htb!]
\includegraphics[ width=3.2in] {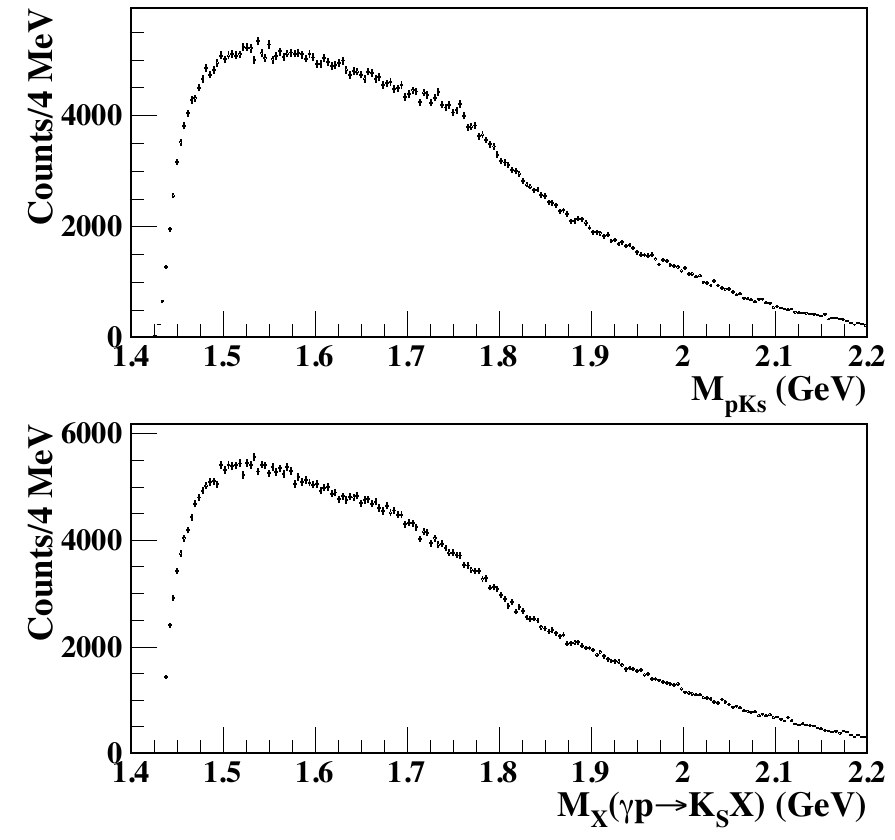}
\caption{(Color online) The  $pK_S$ invariant mass (top) and $K_S$ missing mass (bottom) for the reaction $\gamma p \to \bar K^0 K^0p$ after all cuts,  see ~\cite{CLAS1} for more details. }
\label{fig:batt3}
\end{figure}

In this reaction  there are no $\Lambda^*s$ production  possible,  one could have applied a cut on $M(pK_S)$ and vice versa on $M(pK_L)$ below 1.52~GeV or so, still having enough phase space for the $\Theta^+$ production.  The final plot of $M(pK_S) $ and $M_X(K_S)$ is presented in Fig.~\ref{fig:batt3}. The reason of why $\Theta^+$ is seen in some experiments and not in others was aslo discussed in ref.~\cite{AGS}.
Another review articles from different perspectives are presented in ref.~\cite{hicks} and ~\cite{schumacher}.

\section{Interference}

\begin{figure}[htb!]
\includegraphics[ width=3.2in] {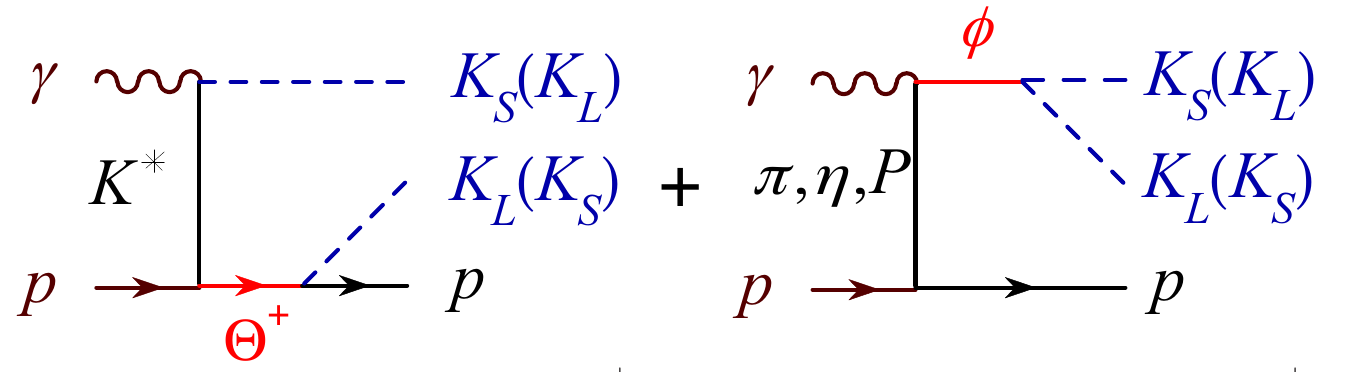}
\caption{(Color online) Two different sub processes that could lead to the same final state in the reaction $\gamma p \to pK_SK_L$. The left diagram is for a $\Theta^+$ production, the right- for a $\phi$ production.}

\label{fig:feynman}
\end{figure}

\begin{figure}[htb!]
\includegraphics[ width=3.2in] {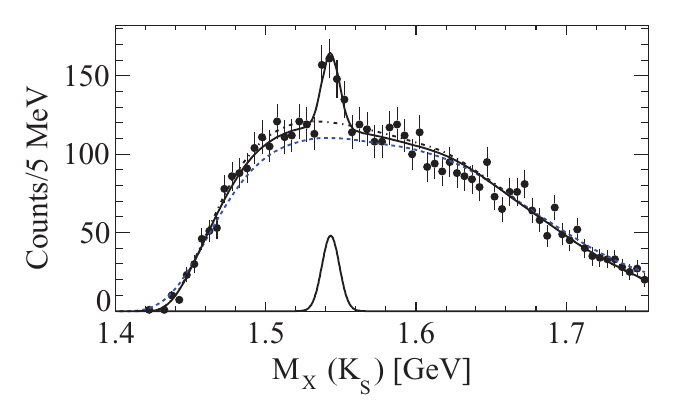}
\caption{(Color online) Missing mass of $K_S$. The dashed line is  the result of the photoproduction of the $\phi$ meson  Monte Carlo. The dashed-dotted line is a modified Monte Carlo distribution, and the solid line is the result of a fit with a modified Monte Carlo distribution plus Gaussian function. }
\label{fig:interference}
\end{figure}

As we learned from previous sections there are two main problems in the searches for $\Theta^+$. One of them is how to suppress reflections in the 3-body (or  4-body) final states in the photoproduction reactions searching for a peak in the invariant mass of either $K^+n$ or $K^0 p$. The second problem is how confidently reproduce  the background.  Both of these problems can be solved if one looks at the interference between two reactions: 
$\gamma p \to p \phi( K_S K_L)$ and $\gamma p \to K_S(K_L) \Theta^+$,  which was proposed in ref.~\cite{amarian}. The feynamn diagrams for these two processes are presented in Fig.~\ref{fig:feynman}.  By selecting events under the $\phi$ peak, we plotted the missing mass of $K_S$ presented in Fig.~\ref{fig:interference}. The only cut we apllied was a cut on a $t$-Mandelstam $-t_{\Theta} <0.45$~GeV$^2$.  The photoproduction of the $\phi$ meson was generated using Monte Carlo (MC) simulation and reconstructed using CLAS reconstruction software and is presented as a dashed line.  In order to take into account possible imperfections in the reconstruction program we allowed the $\phi$ MC distribution to vary presented as a dashed-dotted line in the figure. The solid line is the result of a fit with modified MC for the background plus Gaussian function. The statistical significance of the peak was estimated as log likelihood ratio and was found to be 5.3$\sigma$, corresponding to the ratio of the signal hypothesis over the  background only hypothesis to be about 1.3M.  

Although this analysis 
was under the review for almost five years and we produced thousands of plots per request of the CLAS collaboration members.  At then end the collaboration as a whole didn't sign the paper motivating it because of the narrow kinematic range of the observed signal appearing at  low $t_{\Theta}$.  

In this search
the background model is well under control and low-$t_{\Theta}$ may select the most prominent region for the interference of these two sub processes. The observed peak value is $M_X(K_s)=1.543\pm 0.001$~GeV with a Gaussian width $\sigma=0.004 \pm 0.001$~GeV. For the completeness let us mention the ref.~\cite{azimov},  where quantum mechanical interference was discussed as a powerful tool to observe hadron resonances.

\section{Summary}

As I tried to explain, there is no doubt that in some experiments the statistical significance of the observation of resonance structure that could be associated with purported $\Theta^+$ pentaquark was overestimated. The high statistics experiments performed by the CLAS collaboration showed it dismissing previous CLAS claims.  Nevertheless the analysis of the missing mass of $K_S$  for events under the $\phi$ peak clearly shows that the peak of $\Theta^+$ is observed with statistical significance of 5.3$\sigma$.
Besides these,  as it was discussed,  the performed analyses of high statistics experiments with CLAS were mainly oriented to disprove previous claims and were not used fully to make independent searches by eliminating reflections from other sub processes in the same reactions.  Finally,  let me mention that the most direct way to observe $\Theta^+$ would be by secondary beams of kaons, especially at the approved experiment with K-long facility in Hall D at JLab ~\cite{KLF} in a two-body reaction $K_L p \to K^+n$ with the $M(K^+n)$ resolution on the order of 1-2~MeV in the range from almost the threshold up to 1.6~GeV measured simultaneously because of the broad momentum range of the secondary $K_L$ beam impinging on the hydrogen target,  contrary to the charged kaon beams with the fixed beam energy.  Moreover another exotic member of anti-decuplet, the $\Xi^+$ could be observed in the reaction $K_L p \to K_S\Xi^+$.

In this paper I critically reviewed the experimental searches performed so far to observe the  $\Theta^+$ pentaquark. It is shown that current status of non observation of this particle in some of them  doesn't allow to dismiss its existence as it is commonly accepted.

At the end let me mention that recent results on pentaquarks ~\cite{LHCB}, although in the heavy quark domain,  may have reduced the level of skepticism on the existence of pentaquarks and the Bayesian prior may not be assumed to be zero for the existence of the light pentaquark,  as it seems the majority in the community is inclined to believe.
 
\section*{ACKNOWLEDGMENTS}
I am thankful to  many collegues for over many years of conversations on a different aspects of pentaquarks, namely Dmitri Diakonov, 
Victor Petrov, Maxim Polyakov,  Yakov Azimov, Rober Jaffe, Frank Wilczek,  Gail Dodge,  Charles Hyde,  Gagik Gavalian,  Igor Strakovsky and others.

This work was supported in part  by the U. S. Department of Energy,  Office of Science, Office of Nuclear Physics, under Award Number DE-FG02-96ER40960.


\clearpage
\begin{thebibliography}{99}

\bibitem{gell-mann}
M.~Gell-Mann, Phys. \ Lett. {\bf 8}, 214 (1964).
\bibitem{LHCB} R.~Aaij et. al., [LHCb Collaboration],  Phys.  \ Rev. \ Lett.{\bf 115},  072001, (2015).
\bibitem{DIAKON}
D.~Diakonov, V.~Petrov, and M.V.~Polyakov,
Z.\ Phys.\  A {\bf 359}, 305 (1997)
\bibitem{nakano} T.~Nakano {\it et.  al.}, Phys. \ Rev. \ Lett. {\bf 91}, 012002 (2003).
\bibitem{DIANA} V.V.~Barmin {\it et al.} [DIANA Collaboration], Phys. \ Atom. \ Nucl. {\bf 66}, 1715 (2003); Yad. \ Fiz. {\bf 66}, 1763 (2003).
\bibitem{CLASD} S.~Stepanyan {\it et al.} [CLAS Collaboration], Phys. \ Rev. \ Lett. {\bf 91}, 252001 (2003).
\bibitem{CLASH} V.~Kubarovsky {\it et al.} [CLAS Collaboration], Phys. \ Rev. \ Lett. {\bf 92}, 032001 (2004).
\bibitem{SAPHIR} J.~Barth {\it et al.} [SAPHIR Collaboration], Phys. \ Lett. B {\bf 572}, 127 (2003).
\bibitem{HERMES} A.~Airapetian {\it et al.}, [HERMES Collaboration], Phys. \ Lett. B {\bf 585}, 213 (2004).
\bibitem{ZEUS} S.~Chekanov {\it et al.}, [ZEUS Collaboration], Phys. \ Lett. B {\bf 591}, 7 (2004).
\bibitem{COSY} M.~Abdel-Bary {\it et al.}, [COSY-TOF Collaboration], Phys. \ Lett. B {\bf 595}, 127 (2004). 
\bibitem{SVD} A.~Aleev {\it et al.}, [SVD Collaboration], Yad. \
 Fiz. {\bf 68}, 1012 (2005).
\bibitem{HERAB} I.~Abt, {\it et al.}, [HERA-B Collaboration], Phys. \
Rev. \ Lett. {\bf 93}, 212003 (2004).
\bibitem{HyperCP} M.J.~Longo, {\it et al.}, [HyperCP Collaboration], Phys. \
Rev. D {\bf 70}, 111101 (2004).
\bibitem{BES} J.Z.~Bai, {\it et al.}, [BES Collaboration], Phys. \
Rev. D {\bf 70}, 034506 (2004).
\bibitem{ALEPH} R.~Barate, {\it et al.}, [ALEPH Collaboration], Phys. \ Lett. {\bf 599}, 1 (2004). 
\bibitem{BABAR} B.~Aubert, {\it et al.}, [BABAR Collaboration], Phys. \
Rev. \ Lett. {\bf 95}, 042002 (2005).
 \bibitem{CLASg10} B.~McKinnon {\it et al.} [CLAS Collaboration], Phys. Rev. Lett. {\bf 96}, 212001 (2006).
 \bibitem{CLAS1}
 R.~De Vita {\it et al.}  [CLAS Collaboration],  Phys.\ Rev.\  D {\bf 74}, 032001 (2006)
[arXiv:hep-ex/0606062].
\bibitem{AGS} Y.~Azimov, K.~Goeke and I.~Strakovsky, Phys. \ Rev. D {\bf 76}, 074013 (2007)
[arXiv:0708.2675]
\bibitem{hicks} K.~Hicks,  Eur.  \ Phys.  \ J.  H~{\bf 37}, 1-31 (2012).
\bibitem{schumacher} R.~Schumacher,  [arXiv:nucl-ex/0512042].
\bibitem{amarian}  M.~Amarian, D.~Diakonov, and M.V.~Polyakov,
Phys.\ Rev.\  D {\bf 78}, 074003 (2008).
 \bibitem{amaryan}  M.J.~Amaryan {\it et al.}, Phys.\ Rev.\ C {\bf     85}, 035209 (2012) [arXiv:hep-ex/1110.3325].  
 \bibitem{azimov} Ya.~Azimov, J. \ Phys. \ G: \ Nucl. \ Part. \ Phys. {\bf 37}, 023001 (2010) 023001
 \bibitem{KLF} M.~Amaryan {\it et al.} [KLF Collaboration],  [arXiv:2008.08215].

513 (2003).




\end{thebibliography}
\end{document}